\pdfoutput=1
\documentclass{elsarticle}
\usepackage[T1]{fontenc}
\usepackage[latin9]{inputenc}
\usepackage{units}
\usepackage{mathrsfs}
\usepackage{amsmath}
\usepackage{amssymb}
\usepackage{graphicx}
\usepackage{esint}

\makeatletter

\providecommand{\tabularnewline}{\\}

\makeatother

\begin{document}

\title{Numerical stability analysis of the Pseudo-Spectral Analytical Time-Domain
PIC algorithm}

\author{Brendan B. Godfrey}

\address{University of Maryland, College Park, Maryland 20742, USA}

\author{Jean-Luc Vay}

\address{Lawrence Berkeley National Laboratory, Berkeley, California 94720,
USA}

\author{Irving Haber}

\address{University of Maryland, College Park, Maryland 20742, USA}
\begin{abstract}
The pseudo-spectral analytical time-domain (PSATD) particle-in-cell
(PIC) algorithm solves the vacuum Maxwell's equations exactly, has
no Courant time-step limit (as conventionally defined), and offers
substantial flexibility in plasma and particle beam simulations. It
is, however, not free of the usual numerical instabilities, including
the numerical Cherenkov instability, when applied to relativistic
beam simulations. This paper derives and solves the numerical dispersion
relation for the PSATD algorithm and compares the results with corresponding
behavior of the more conventional pseudo-spectral time-domain (PSTD)
and finite difference time-domain (FDTD) algorithms. In general, PSATD
offers superior stability properties over a reasonable range of time
steps. More importantly, one version of the PSATD algorithm, when
combined with digital filtering, is almost completely free of the
numerical Cherenkov instability for time steps (scaled to the speed
of light) comparable to or smaller than the axial cell size.\end{abstract}
\maketitle

\section{Introduction}

Particle in Cell (PIC) plasma simulation codes typically employ a
Finite Difference Time Domain (FDTD) algorithm with staggered spatial
mesh \citep{Yee} for advancing Maxwell's equations. The FDTD algorithm
is straightforward, second-order accurate, and parallelizes well for
efficient computation on modern computers. A very flexible and at
least equally accurate approach for solving Maxwell's equations numerically
is the pseudo-spectral method, in which Maxwell's equations are Fourier-decomposed
in space, and the resulting equations advanced in time to second-order
or better accuracy \citep{Liu1997}. Pseudo-spectral PIC algorithms
require Fourier-transforming the currents and fields at every time-step,
because particles are advanced in real rather than Fourier space.
Because Fourier transforms generally do not parallelize well, pseudo-spectral
methods are used less commonly than FDTD methods in PIC codes.

Nonetheless, the advantages of pseudo-spectral methods should not
be ignored. Haber's Pseudo-Spectral Analytical Time Domain (PSATD)
algorithm \citep{HaberICNSP73}, in particular, is exact for plasma
currents constant in time and, consequently, is free of electromagnetic
wave numerical dispersion for wave numbers satisfying $k\leq\pi/\triangle t$
and has no Courant limit in the usual sense. It also offers highly
accurate balancing of the Lorentz force, $\mathbf{E}+\mathbf{v}\times\mathbf{B}$
\citep{Vay2008}, which is especially desirable in simulations of
relativistic beams or of Laser-Plasma Acceleration (LPA) in frames
co-moving with the interaction region \citep{VayPRL07,VayPoP2011}.
PSATD also has superior numerical stability properties \citep{Vay2013PSATD}.
The more commonly used Pseudo-Spectral Time Domain (PSTD) algorithm
\citep{Liu1997,Xu2012} enjoys some of these same advantages but has
a restrictive Courant limit.

Importantly, a domain decomposition method recently has been developed
that allows efficient parallelization of Fourier transforms \citep{Vay2013PSATD}
in PIC codes. It takes advantage of the linearity and finite propagation
velocity of light in Maxwell\textquoteright{}s equations to limit
communication of data between neighboring computational domains. The
small approximation required appears to be insignificant for a range
of problems of interest.

Despite the advantages of pseudo-spectral methods, they are known
not to be free of the numerical Cherenkov instability \citep{godfrey1974numerical},
which results from coupling of electromagnetic waves with numerically
spurious beam mode aliases in cold beam simulations. In this paper,
the numerical dispersion relation is derived for the PSATD algorithm
with either a version of the Esirkepov algorithm \citep{esirkepov2001exact}
or conventional current interpolation. Although the PSATD algorithm
does not exhibit special time-steps at which numerical instability
growth rates are very small \citep{Xu2012,VayJCP2011,godfrey2013esirkepov},
a slight generalization of the PSATD-Esirkepov combination is shown
to have extraordinarily good stability properties when cubic interpolation
and appropriate digital filtering are employed, certainly substantially
better than that of FDTD algorithms previously analyzed \citep{godfrey2013esirkepov}.
The PSTD-Esirkepov algorithm also has good stability properties over
its range of allowed time-steps, although not quite as good as that
of the PSATD-Esirkepov algorithm. These analyses have been confirmed
using the multidimensional WARP \citep{Warp} PIC code for two-dimensional
simulations of plasma wake formation in a LPA stage. The parameters
used for the WARP simulations were similar to those used in \citep{godfrey2013esirkepov}.
However, the length of the plasma was increased thirty-fold, due to
the extremely small growth rates that were observed when using the
PSATD solver.

The remainder of this paper is organized as follows. The PSATD algorithm
coupled with either the Esirkepov or the conventional current deposition
algorithm is presented in Sec. 2. Derivations of the corresponding
numerical instability dispersion relations for multidimensional PSATD
PIC codes are outlined briefly in Sec. 3. The dispersion relations
are specialized in Sec. 4 to a cold, relativistic beam in two dimensions
for comparison with WARP simulations. Sec. 5 provides a reasonably
accurate approximation for maximum numerical instability growth rates
for the PSATD-Esirkepov algorithm with digital filtering, showing
the desirable numerical stability properties just mentioned. Then,
the dispersion relations are solved numerically for a range of options
and parameters and compared with WARP results in Sec. 6. (These analytical
and numerical dispersion relation calculations were performed using
\emph{Mathematica} \citep{Mathematica9}.) As a comparison, stability
results for the more commonly used PSTD algorithm are derived and
discussed in Sec. 6. Sec. 7 presents WARP simulations, demonstrating
the near absence of numerical instabilities in actual LPA simulations
for appropriately chosen options and time-steps. The concluding section
summarizes the findings in the paper and compares them with corresponding
FDTD results.

\section{PSATD algorithm}

The PSATD algorithm is derived in some detail in Appendix A of \citep{Vay2013PSATD}
and presented in Eqs. (13) and (14) of that article. It also can be
obtained directly by integrating analytically the spatially Fourier-transformed
Maxwell's equations, Eqs. (1) and (2) of \citep{Vay2013PSATD}, for
one time-step under the assumption that currents are constant over
the time-step. In either case, the algorithm is
\begin{multline}
\mathbf{E}^{n+1}=C\mathbf{E}^{n}-iS\mathbf{k}\times\mathbf{B}^{n}/k-S\mathbf{J}^{n+\nicefrac{1}{2}}/k+\left(1-C\right)\mathbf{k}\mathbf{k}\cdot\mathbf{E}^{n}/k^{2}\\
+\left(S/k-\triangle t\right)\mathbf{k}\mathbf{k}\cdot\mathbf{J}^{n+\nicefrac{1}{2}}/k^{2},\label{eq:PSATD-E}
\end{multline}
\begin{equation}
\mathbf{B}^{n+1}=C\mathbf{B}^{n}+iS\mathbf{k}\times\mathbf{E}^{n}/k-i\left(1-C\right)\mathbf{k}\times\mathbf{J}^{n+\nicefrac{1}{2}}/k^{2},\label{eq:PSATD-B}
\end{equation}
with $\mathbf{k}$ the wave-number, $k$ its magnitude, $C=\cos\left(k\triangle t\right)$,
and $S=\sin\left(k\triangle t\right)$. The speed of light is normalized
to unity. Note that the sign of $\mathbf{k}$ is reversed relative
to \citep{Vay2013PSATD} for consistency with earlier analyses of
the numerical Cherenkov instability, \emph{e.g.}, \citep{godfrey2013esirkepov,godfrey1975canonical}. 

Eqs. (\ref{eq:PSATD-E}) and (\ref{eq:PSATD-B}) define both $\mathbf{E}^{n}$
and $\mathbf{B}^{n}$ at integer time-steps. For deriving the PSATD
numerical dispersion relation, and perhaps also for implementing the
PSATD algorithm in some PIC simulation codes, a leap-frog arrangement
in which $\mathbf{B}$ is defined at half-integer time-steps is more
convenient. To do so, we simply define $\mathbf{B}^{n+\nicefrac{1}{2}}$
at half-integer time-steps as 
\begin{equation}
\mathbf{B}^{n}=\frac{1}{2C_{h}}\left(\mathbf{B}^{n+\nicefrac{1}{2}}+\mathbf{B}^{n-\nicefrac{1}{2}}\right).\label{eq:Bave}
\end{equation}
Using this equation, we can eliminate $\mathbf{B}^{n}$ at integer
time-steps from Eqs. (\ref{eq:PSATD-E}) and (\ref{eq:PSATD-B}) to
obtain
\begin{equation}
\mathbf{E}^{n+1}=\mathbf{E}^{n}-2iS_{h}\mathbf{k}\times\mathbf{B}^{n+\nicefrac{1}{2}}/k-S\mathbf{J}^{n+\nicefrac{1}{2}}/k+\left(S/k-\Delta t\right)\mathbf{k}\mathbf{k}\cdot\mathbf{J}^{n+\nicefrac{1}{2}}/k^{2},\label{eq:leapfrogE}
\end{equation}
\begin{equation}
\mathbf{B}^{n+\nicefrac{3}{2}}=\mathbf{B}^{n+\nicefrac{1}{2}}+2iS_{h}\mathbf{k}\times\mathbf{E}^{n+1}/k,\label{eq:leapfrogB}
\end{equation}
after a modest amount of algebra. Here, $C_{h}=\cos\left(k\triangle t/2\right)$,
and $S_{h}=\sin\left(k\triangle t/2\right)$. (Note that Eqs. (\ref{eq:leapfrogE})
and (\ref{eq:leapfrogB}) differ from Eqs. (15) and (16) of \citep{Vay2013PSATD},
which are based on a different definition of $\mathbf{B}^{n+\nicefrac{1}{2}}$.)

The divergence of Eq. (\ref{eq:leapfrogE}) yields $\mathbf{k}\cdot\mathbf{E}^{n+1}=\mathbf{k}\cdot\mathbf{E}^{n}-\mathbf{k}\cdot\mathbf{J}^{n+\nicefrac{1}{2}}\triangle t,$
which assures that $\mathbf{k}\cdot\mathbf{E}^{n+1}=i\rho^{n+1}$,
provided that charge is conserved, 
\begin{equation}
\mathbf{k}\cdot\mathbf{J}^{n+\nicefrac{1}{2}}=-i\left(\rho^{n+1}-\rho^{n}\right)/\triangle t\label{eq:continuity}
\end{equation}
(and also provided that $\mathbf{k}\cdot\mathbf{E}^{0}=i\rho^{0}$
at initialization). The Buneman current deposition algorithm \citep{VillasenorCPC92}
and its generalization, the Esirkepov algorithm \citep{esirkepov2001exact},
satisfy the discretized continuity equation in real space. The adaptation
of the Esikepov algorithm for k-space in Eq. (20) of \citep{Vay2013PSATD}
automatically satisfies Eq. (\ref{eq:continuity}). (This modification
of the Esirekpov algorithm for PSATD will be referred to as the Esirkepovk
algorithm in the remainder of the paper.) Otherwise, (\ref{eq:leapfrogE})
must be rewritten as
\begin{multline}
\mathbf{E}^{n+1}=\mathbf{E}^{n}-2iS_{h}\mathbf{k}\times\mathbf{B}^{n+\nicefrac{1}{2}}/k-S\mathbf{J}^{n+\nicefrac{1}{2}}/k\\
+S\mathbf{k}\mathbf{k}\cdot\mathbf{J}^{n+\nicefrac{1}{2}}/k^{3}+i\mathbf{k}\left(\rho^{n+1}-\rho^{n}\right)/k^{2},\label{eq:leapfrogE-alt}
\end{multline}
which has as its divergence, $\mathbf{k}\cdot\mathbf{E}^{n+1}=\mathbf{k}\cdot\mathbf{E}^{n}+i\left(\rho^{n+1}-\rho^{n}\right),$
as desired. In subsequent sections both charge-conserving and non-charge-conserving
PSATD variants will be analyzed, and Eq. (\ref{eq:leapfrogE-alt})
will be used instead of Eq. (\ref{eq:leapfrogE}) in the latter instances.

As we shall see in Sec. 6, scaling the Esirkepovk currents by k-dependent
factors $\zeta$ can be beneficial for numerical stability; i.e.,
$\mathbf{J}=\mathbf{\mathbf{\zeta}:}\mathbf{J}_{e}$, with $\mathbf{\zeta}=$diag$\left(\zeta_{z},\zeta_{x},\zeta_{y}\right)$
and $\mathbf{J}_{e}$ the current computed by the Esirkepovk algorithm.
Doing so, of course, requires the use of Eq. \ref{eq:leapfrogE-alt},
because introducing the factors $\zeta$ typically does not preserve
charge conservation. However, because the Esirkepovk current satisfies
Eq. \ref{eq:continuity} identically, Eq. \ref{eq:leapfrogE-alt}
can be rewritten in this case as
\begin{multline}
\mathbf{E}^{n+1}=\mathbf{E}^{n}-2iS_{h}\mathbf{k}\times\mathbf{B}^{n+\nicefrac{1}{2}}/k-S\mathbf{\mathbf{\zeta}:}\mathbf{J}_{e}^{n+\nicefrac{1}{2}}/k\\
+S\mathbf{k}\mathbf{k}\cdot\mathbf{\mathbf{\zeta}:}\mathbf{J}_{e}^{n+\nicefrac{1}{2}}/k^{3}-\mathbf{k}\mathbf{k}\cdot\mathbf{J}_{e}^{n+\nicefrac{1}{2}}\triangle t/k^{2},\label{eq:leapfrog-alt1}
\end{multline}
which can be viewed as a generalization of Eq. \ref{eq:leapfrogE}.

The divergence of (\ref{eq:leapfrogB}) yields $\mathbf{k}\cdot\mathbf{B}^{n+\nicefrac{3}{2}}=\mathbf{k}\cdot\mathbf{B}^{n+\nicefrac{1}{2}}$,
assuring that $\mathbf{k}\cdot\mathbf{B}^{n+\nicefrac{3}{2}}=0$,
if it is so at initialization.

\section{Numerical instability dispersion relation}

The derivation of the numerical instability dispersion relation for
the PSATD and Esirkepovk combined algorithm follows closely the corresponding
derivation for the FDTD and Esirkepov combined algorithm in \citep{godfrey2013esirkepov}.
To begin, the temporal Fourier transforms of Eqs. (\ref{eq:leapfrog-alt1})
and (\ref{eq:leapfrogB}) are
\begin{equation}
\left[\omega\right]\mathbf{E}=-2S_{h}\mathbf{k}\times\mathbf{B}/k+iS\mathbf{\mathbf{\zeta}:}\mathbf{J}_{e}/k-iS\mathbf{k}\mathbf{k}\cdot\mathbf{\mathbf{\zeta}:}\mathbf{J}_{e}/k^{3}+i\mathbf{k}\mathbf{k}\cdot\mathbf{J}_{e}\Delta t/k^{2},\label{eq:Etrans}
\end{equation}
\begin{equation}
\left[\omega\right]\mathbf{B}=2S_{h}\mathbf{k}\times\mathbf{E}/k.\label{eq:Btrans}
\end{equation}
Brackets around the frequency, $\omega$, designate its finite difference
(leapfrog) representation,
\begin{equation}
\left[\omega\right]=\sin\left(\omega\frac{\Delta t}{2}\right)/\left(\frac{\Delta t}{2}\right).\label{eq:meshom}
\end{equation}

The Esirkepov algorithm, either in real or k-space, determines not
the current itself but its first derivative \citep{esirkepov2001exact}.
In the PSATD algorithm, that derivative is given by $\mathbf{k}$,
not $\left[\mathbf{k}\right]$. Consequently, Eq. (5) of \citep{godfrey2013esirkepov}
becomes
\begin{equation}
\left\{ \begin{array}{c}
W_{x}\\
W_{y}\\
W_{z}
\end{array}\right\} =-i\Delta t\left\{ \begin{array}{c}
k_{x}\mathscr{\mathcal{J}}_{x}\\
k_{y}\mathcal{J}_{y}\\
k_{z}\mathcal{J}_{z}
\end{array}\right\} ,\label{eq:dJ}
\end{equation}
and the current contribution from an individual particle, Eq. (7)
of \citep{godfrey2013esirkepov}, becomes
\begin{equation}
\left\{ \begin{array}{c}
\mathcal{J}_{x}\\
\mathcal{J}_{y}\\
\mathcal{J}_{z}
\end{array}\right\} =S^{J}\frac{2}{\Delta t}\left\{ \begin{array}{c}
\sin\left(k_{x}^{\prime}v_{x}\frac{\Delta t}{2}\right)\left[\cos\left(k_{y}^{\prime}v_{y}\frac{\Delta t}{2}\right)\cos\left(k_{z}^{\prime}v_{z}\frac{\Delta t}{2}\right)-\frac{1}{3}\sin\left(k_{y}^{\prime}v_{y}\frac{\Delta t}{2}\right)\sin\left(k_{z}^{\prime}v_{z}\frac{\Delta t}{2}\right)\right]/k_{x}\\
\sin\left(k_{y}^{\prime}v_{y}\frac{\Delta t}{2}\right)\left[\cos\left(k_{z}^{\prime}v_{z}\frac{\Delta t}{2}\right)\cos\left(k_{x}^{\prime}v_{x}\frac{\Delta t}{2}\right)-\frac{1}{3}\sin\left(k_{z}^{\prime}v_{z}\frac{\Delta t}{2}\right)\sin\left(k_{x}^{\prime}v_{x}\frac{\Delta t}{2}\right)\right]/k_{y}\\
\sin\left(k_{z}^{\prime}v_{z}\frac{\Delta t}{2}\right)\left[\cos\left(k_{x}^{\prime}x_{y}\frac{\Delta t}{2}\right)\cos\left(k_{y}^{\prime}v_{y}\frac{\Delta t}{2}\right)-\frac{1}{3}\sin\left(k_{x}^{\prime}v_{x}\frac{\Delta t}{2}\right)\sin\left(k_{y}^{\prime}v_{y}\frac{\Delta t}{2}\right)\right]/k_{z}
\end{array}\right\} ,\label{eq:Jpart}
\end{equation}
with $S^{J}$ the current interpolation function. Finally, the total
current is given by Eq. (10) of \citep{godfrey2013esirkepov}, 
\begin{equation}
\mathbf{J}=\sum_{m}\int\mathbf{F\cdot\frac{\partial}{\partial\mathbf{p}}\,\mathcal{\boldsymbol{J}}\,\csc}\left[\left(\omega-\mathbf{k^{\prime}\cdot v}\right)\frac{\Delta t}{2}\right]\frac{\Delta t}{2}f\,\mathrm{d}^{3}\mathbf{v},\label{eq:J}
\end{equation}
summed over spatial aliases. The determinant of the 6x6 matrix comprised
of Eqs. (\ref{eq:Etrans}), (\ref{eq:Btrans}), and (\ref{eq:J})
is the desired PSATD-Esirkepovk dispersion relation. 

Alternatively, the current can be accumulated at nodal points by conventional
interpolation, in which case charge is not conserved automatically,
and Eq. \ref{eq:leapfrogE-alt} should be used. Its temporal Fourier
transform is
\begin{equation}
\left[\omega\right]\mathbf{E}=-2S_{h}\mathbf{k}\times\mathbf{B}/k+iS\mathbf{J}/k-iS\mathbf{k}\mathbf{k}\cdot\mathbf{J}/k^{3}+i\left[\omega\right]\mathbf{k}\rho/k^{2}.\label{eq:Etrans-alt}
\end{equation}
 Currents are interpolated directly to nodes on the grid, so Eq. (\ref{eq:J})
becomes
\begin{equation}
\mathbf{J}=\sum_{m}S^{J}\int\mathbf{F\cdot\frac{\partial}{\partial\mathbf{p}}\,\mathbf{v}\,\csc}\left[\left(\omega-\mathbf{k^{\prime}\cdot v}\right)\frac{\Delta t}{2}\right]\frac{\Delta t}{2}f\,\mathrm{d}^{3}\mathbf{v}.\label{eq:J-alt}
\end{equation}
Similarly, the charge density is given by \citep{godfrey1975canonical}
\begin{equation}
\mathbf{\rho}=\sum_{m}S^{J}\int\mathbf{F\cdot\frac{\partial}{\partial\mathbf{p}}\,\cot}\left[\left(\omega-\mathbf{k^{\prime}\cdot v}\right)\frac{\Delta t}{2}\right]\frac{\Delta t}{2}f\,\mathrm{d}^{3}\mathbf{v}.\label{eq:rho}
\end{equation}
(The charge and current interpolation functions are assumed to be
the same.) The dispersion relation in this case is the determinant
of the 6x6 matrix comprised of Eqs. (\ref{eq:Etrans-alt}), (\ref{eq:Btrans}),
(\ref{eq:J-alt}), and (\ref{eq:rho}).

\section{WARP-PSATD 2-d dispersion relation}

For comparison with WARP-PSATD-Esirkepovk two-dimensional, cold beam
simulation results, we reduce Eqs. (\ref{eq:Etrans}) and (\ref{eq:Btrans})
to a 3x3 system in $\left\{ E_{z},E_{x},B_{y}\right\} $ and perform
the integral in Eq. (\ref{eq:J}) for a cold beam propagating at velocity
\textit{v} in the \textit{z}-direction. The resulting matrix equation
is
\begin{equation}
\left(\begin{array}{ccc}
\xi_{z,z}+[\omega] & \xi_{z,x} & \xi_{z,y}+[k_{x}]\\
\xi_{x,z} & \xi_{x,x}+[\omega] & \xi_{x,y}-[k_{z}]\\
{}[k_{x}] & -[k_{z}] & [\omega]
\end{array}\right)\left(\begin{array}{c}
E_{z}\\
E_{x}\\
B_{y}
\end{array}\right)=0.\label{eq:M3x3}
\end{equation}
Its determinant set equal to zero, 
\begin{multline}
[\omega]\left([\omega]^{2}-[k_{z}]^{2}-[k_{x}]^{2}\right)+\left([\omega]^{2}-[k_{z}]^{2}\right)\xi_{z,z}-[k_{z}][k_{x}]\xi_{x,z}\\
-[k_{z}][k_{x}]\xi_{z,x}-[\omega][k_{x}]\xi_{z,y}+\left([\omega]^{2}-[k_{x}]^{2}\right)\xi_{x,x}+[\omega][k_{z}]\xi_{x,y}\\
+\xi_{z,z}\left([\omega]\xi_{x,x}+[k_{x}]\xi_{x,y}\right)-\xi_{x,z}\left([\omega]\xi_{z,x}+[k_{z}]\xi_{z,y}\right)\\
+[k_{x}]\left(\xi_{z,x}\xi_{x,y}-\xi_{z,y}\xi_{x,x}\right)=0,\label{eq:det}
\end{multline}
is the dispersion relation. The quantities $\left[\mathbf{k}\right]$
and $\xi$ are introduced purely for notational simplicity.
\begin{equation}
\left[k_{z}\right]=k_{z}\sin\left(k\frac{\Delta t}{2}\right)/\left(k\frac{\Delta t}{2}\right),\label{eq:meshkz}
\end{equation}
\begin{equation}
\left[k_{x}\right]=k_{x}\sin\left(k\frac{\Delta t}{2}\right)/\left(k\frac{\Delta t}{2}\right).\label{eq:meshkx}
\end{equation}
\begin{multline}
\xi_{z,z}=-n\gamma^{-2}\sum_{m}S^{J}S^{E_{z}}\csc^{2}\left[\left(\omega-k_{z}^{\prime}v\right)\frac{\Delta t}{2}\right]\\
\left(kk_{z}^{2}\Delta t+\zeta_{z}k_{x}^{2}\sin\left(k\Delta t\right)\right)\Delta t\left[\omega\right]k_{z}^{\prime}/4k^{3}k_{z}\mathrm{,}\label{eq:Mzz}
\end{multline}
\begin{equation}
\xi_{z,x}=-n\sum_{m}S^{J}S^{E_{x}}\csc\left[\left(\omega-k_{z}^{\prime}v\right)\frac{\Delta t}{2}\right]\eta_{z}k_{x}^{\prime}/2k^{3}k_{z},\label{eq:Mzx}
\end{equation}
\begin{equation}
\xi_{z,y}=nv\sum_{m}S^{J}S^{B_{y}}\csc\left[\left(\omega-k_{z}^{\prime}v\right)\frac{\Delta t}{2}\right]\eta_{z}k_{x}^{\prime}/2k^{3}k_{z},\label{eq:Mzy}
\end{equation}
\begin{multline}
\xi_{x,z}=-n\gamma^{-2}\sum_{m}S^{J}S^{E_{z}}\csc^{2}\left[\left(\omega-k_{z}^{\prime}v\right)\frac{\Delta t}{2}\right]\\
\left(k\Delta t-\zeta_{z}\sin\left(k\Delta t\right)\right)\Delta t\left[\omega\right]k_{x}k_{z}^{\prime}/4k^{3},\label{eq:Mxz}
\end{multline}
\begin{equation}
\xi_{x,x}=-n\sum_{m}S^{J}S^{E_{x}}\csc\left[\left(\omega-k_{z}^{\prime}v\right)\frac{\Delta t}{2}\right]\eta_{x}k_{x}^{\prime}/2k^{3}k_{x},\label{eq:Mxx}
\end{equation}
\begin{equation}
\xi_{x,y}=nv\sum_{m}S^{J}S^{B_{y}}\csc\left[\left(\omega-k_{z}^{\prime}v\right)\frac{\Delta t}{2}\right]\eta_{x}k_{x}^{\prime}/2k^{3}k_{x},\label{eq:Mxy}
\end{equation}
with
\begin{multline}
\eta_{z}=\cot\left[\left(\omega-k_{z}^{\prime}v\right)\frac{\Delta t}{2}\right]\left(kk_{z}^{2}\Delta t+\zeta_{z}k_{x}^{2}\sin\left(k\Delta t\right)\right)\sin\left(k_{z}^{\prime}v\frac{\Delta t}{2}\right)\\
+\left(k\Delta t-\zeta_{x}\sin\left(k\Delta t\right)\right)k_{z}^{2}\cos\left(k_{z}^{\prime}v\frac{\Delta t}{2}\right),\label{eq:etaz}
\end{multline}
\begin{multline}
\eta_{x}=\cot\left[\left(\omega-k_{z}^{\prime}v\right)\frac{\Delta t}{2}\right]\left(k\Delta t-\zeta_{z}\sin\left(k\Delta t\right)\right)k_{x}^{2}\sin\left(k_{z}^{\prime}v\frac{\Delta t}{2}\right)\\
+\left(kk_{x}^{2}\Delta t+\zeta_{x}k_{z}^{2}\sin\left(k\Delta t\right)\right)\cos\left(k_{z}^{\prime}v\frac{\Delta t}{2}\right).\label{eq:etax}
\end{multline}
Sums are over spatial aliases, $k_{z}^{\prime}=k_{z}+m_{z}\,2\pi/\Delta z$
and $k_{x}^{\prime}=k_{x}+m_{x}\,2\pi/\Delta x$, with $m_{z}$ and
$m_{x}$ integers. $n$ is the beam charge density divided by $\gamma$,
which can be normalized to unity. However, explicitly retaining it
in the dispersion relation sometimes is informative. Eqs. (\ref{eq:Mzz})
- (\ref{eq:etax}) are substantially more complicated than their counterparts
in \citep{godfrey2013esirkepov}, the additional terms arising from
the final expression in Eq. (\ref{eq:PSATD-E}).

Like most other PIC codes, WARP employs splines for current and field
interpolation. The Fourier transform of the current interpolation
function is
\begin{equation}
S^{J}=\left[\sin\left(k_{z}^{\prime}\frac{\Delta z}{2}\right)/\left(k_{z}^{\prime}\frac{\Delta z}{2}\right)\right]^{\ell_{z}+1}\left[\sin\left(k_{x}^{\prime}\frac{\Delta x}{2}\right)/\left(k_{x}^{\prime}\frac{\Delta x}{2}\right)\right]^{\ell_{x}+1};\label{eq:SJ}
\end{equation}
$\ell_{z}$ and $\ell_{x}$ are the orders of the current interpolation
splines in the z- and x-directions. Fields typically are interpolated
with splines of the same centering and order in PSATD implementations,
so $S^{E_{z}}=S^{E_{x}}=S^{J}$. The magnetic field interpolation
function also includes the conversion factor from $\mathbf{B}$ at
half-integer time-steps, as given in Eq. (\ref{eq:leapfrogB}), to
$\mathbf{B}$ at integer time-steps, as used to push the particles.
Hence, from the temporal Fourier transform of Eq. (\ref{eq:Bave}),
$S^{B_{y}}=S^{J}\cos\left(\omega\,\Delta t/2\right)/\cos\left(k\,\Delta t/2\right)$.
With $S^{E_{x}}$ and $S^{B_{y}}$ as just define,
\begin{equation}
\xi_{z,y}/\xi_{z,x}=\xi_{x,y}/\xi_{x,x}=-v\cos\left(\omega\,\Delta t/2\right)/\cos\left(k\,\Delta t/2\right)\label{eq:Mratio}
\end{equation}

Note, however, that field interpolation from a staggered mesh could
be employed instead, as it is in the FDTD version of WARP and most
other PIC codes. In that case the field interpolation functions would
be as described in Eqs. (21) - (23) of \citep{godfrey2013esirkepov}
and the associated text. Eq. (\ref{eq:Mratio}) is only approximately
satisfied for staggered mesh interpolation.

Next, we present the dispersion matrix for WARP-PSATD with conventional
current (and charge) deposition, as described in the final paragraph
of Sec. 3.
\begin{multline}
\xi_{z,z}=-n\gamma^{-2}\sum_{m}S^{J}S^{E_{z}}\csc^{2}\left[\left(\omega-k_{z}^{\prime}v\right)\frac{\Delta t}{2}\right]\Delta t\\
\left\{ \left(\sin\left[\left(\omega-k_{z}^{\prime}v\right)\frac{\Delta t}{2}\right]k_{z}^{\prime}v+\frac{2}{\Delta t}\right)k_{x}^{2}\sin\left(k_{z}\Delta t\right)+k\, k_{z}k_{z}^{\prime}\left[\omega\right]\Delta t\right\} /4k^{3},\label{eq:Mzz-alt}
\end{multline}
\begin{multline}
\xi_{z,x}=-n\sum_{m}S^{J}S^{E_{x}}\csc^{2}\left[\left(\omega-k_{z}^{\prime}v\right)\frac{\Delta t}{2}\right]\Delta t\\
\left\{ \left(\sin\left[\left(\omega-k_{z}^{\prime}v\right)\frac{\Delta t}{2}\right]k_{x}k_{x}^{\prime}v-\frac{2}{\Delta t}k_{z}\right)k_{x}\sin\left(k_{z}\Delta t\right)+k\, k_{z}k_{x}^{\prime}\left[\omega\right]\Delta t\right\} /4k^{3},\label{eq:Mzx-alt}
\end{multline}
\begin{multline}
\xi_{z,y}=nv\sum_{m}S^{J}S^{B_{y}}\csc^{2}\left[\left(\omega-k_{z}^{\prime}v\right)\frac{\Delta t}{2}\right]\Delta t\\
\left\{ \left(\sin\left[\left(\omega-k_{z}^{\prime}v\right)\frac{\Delta t}{2}\right]k_{x}k_{x}^{\prime}v-\frac{2}{\Delta t}k_{z}\right)k_{x}\sin\left(k_{z}\Delta t\right)+k\, k_{z}k_{x}^{\prime}\left[\omega\right]\Delta t\right\} /4k^{3},\label{eq:Mzy-alt}
\end{multline}
\begin{multline}
\xi_{x,z}=-n\gamma^{-2}\sum_{m}S^{J}S^{E_{z}}\csc^{2}\left[\left(\omega-k_{z}^{\prime}v\right)\frac{\Delta t}{2}\right]\Delta t\\
\left\{ \left(\sin\left[\left(\omega-k_{z}^{\prime}v\right)\frac{\Delta t}{2}\right]k_{z}^{\prime}v+\frac{2}{\Delta t}\right)k_{z}\sin\left(k_{z}\Delta t\right)+k\, k_{z}^{\prime}\left[\omega\right]\Delta t\right\} k_{x}/4k^{3},\label{eq:Mxz-alt}
\end{multline}
\begin{multline}
\xi_{x,x}=n\sum_{m}S^{J}S^{E_{x}}\csc\left[\left(\omega-k_{z}^{\prime}v\right)\frac{\Delta t}{2}\right]\Delta t\\
\left\{ \left(\sin\left[\left(\omega-k_{z}^{\prime}v\right)\frac{\Delta t}{2}\right]k_{x}k_{x}^{\prime}v-\frac{2}{\Delta t}k_{z}\right)k_{z}\sin\left(k_{z}\Delta t\right)-k\, k_{x}k_{x}^{\prime}\left[\omega\right]\Delta t\right\} /4k^{3},\label{eq:Mxx-alt}
\end{multline}
\begin{multline}
\xi_{x,y}=-nv\sum_{m}S^{J}S^{B_{y}}\csc\left[\left(\omega-k_{z}^{\prime}v\right)\frac{\Delta t}{2}\right]\Delta t\\
\left\{ \left(\sin\left[\left(\omega-k_{z}^{\prime}v\right)\frac{\Delta t}{2}\right]k_{x}k_{x}^{\prime}v-\frac{2}{\Delta t}k_{z}\right)k_{z}\sin\left(k_{z}\Delta t\right)-k\, k_{x}k_{x}^{\prime}\left[\omega\right]\Delta t\right\} /4k^{3}.\label{eq:Mxy-alt}
\end{multline}
Here too, Eq. (\ref{eq:Mratio}) is satisfied, provided that currents
and fields all are interpolated to and from the same mesh nodes.

As pointed out in \citep{godfrey2013esirkepov}, $m_{x}$ alias terms
in the dispersion relations can be summed explicitly by means of Eqs.
(1.421.3) and (1.422.3) of \citep{GradshteynRyzhik} or derivatives
thereof.

\section{Approximate growth rates}

Useful results can be obtained from the dispersion relation without
solving it in its entirety.

When Eq, (\ref{eq:Mratio}) is satisfied (and approximately otherwise),
\begin{equation}
\xi_{z,x}\xi_{x,y}-\xi_{z,y}\xi_{z,y}=0,\label{eq:minor}
\end{equation}
and the dispersion relation, Eq. \ref{eq:det}, reduces to
\begin{multline}
C_{0}+n\sum_{m_{z}}C_{1}\csc\left[\left(\omega-k_{z}^{\prime}v\right)\frac{\Delta t}{2}\right]+n\sum_{m_{z}}\left(C_{2x}+\gamma^{-2}C_{2z}\right)\csc^{2}\left[\left(\omega-k_{z}^{\prime}v\right)\frac{\Delta t}{2}\right]\\
+\gamma^{-2}n^{2}\left(\sum_{m_{z}}C_{3z}\csc^{2}\left[\left(\omega-k_{z}^{\prime}v\right)\frac{\Delta t}{2}\right]\right)\left(\sum_{m_{z}}C_{3x}\csc\left[\left(\omega-k_{z}^{\prime}v\right)\frac{\Delta t}{2}\right]\right)=0,\label{eq:drformfull}
\end{multline}
with $C_{0}$ the vacuum dispersion function,
\begin{equation}
C_{0}=\left[\omega\right]^{2}-\left[k_{x}\right]^{2}-\left[k_{z}\right]^{2},\label{eq:C0}
\end{equation}
and, for the PSATD-Esirkepovk algorithm,
\begin{multline}
C_{1}=-\sum_{m_{x}}k_{x}^{\prime}\left(S^{J}\right)^{2}\cos\left(k_{z}^{\prime}v\frac{\Delta t}{2}\right)\\
\left\{ \zeta_{x}k_{z}\sin\left(k\,\Delta t\right)\left(k_{z}\sin\left(\omega\frac{\Delta t}{2}\right)-k\, v\tan\left(k\frac{\Delta t}{2}\right)\cos\left(\omega\frac{\Delta t}{2}\right)\right)+k\, k_{x}^{2}\Delta t\, C_{0}/\sin\left(\omega\frac{\Delta t}{2}\right)\right\} /k^{3}k_{x}\Delta t,\label{eq:C1}
\end{multline}
\begin{multline}
C_{2x}=k_{x}\sum_{m_{x}}k_{x}^{\prime}\left(S^{J}\right)^{2}\cos\left[\left(\omega-k_{z}^{\prime}v\right)\frac{\Delta t}{2}\right]\sin\left(k_{z}^{\prime}v\frac{\Delta t}{2}\right)\\
\left\{ \zeta_{z}\sin\left(k\,\Delta t\right)\left(k_{z}\sin\left(\omega\frac{\Delta t}{2}\right)-k\, v\tan\left(k\frac{\Delta t}{2}\right)\cos\left(\omega\frac{\Delta t}{2}\right)\right)-k\, k_{z}\Delta t\, C_{0}/\sin\left(\omega\frac{\Delta t}{2}\right)\right\} /k^{3}k_{z}\Delta t,\label{eq:C2x}
\end{multline}
\begin{multline}
C_{3x}=\sum_{m_{x}}k_{x}^{\prime}\left(S^{J}\right)^{2}\sin\left(k\,\Delta t\right)\cos\left(k_{z}^{\prime}v\frac{\Delta t}{2}\right)\\
\left(k\,\sin\left(\omega\frac{\Delta t}{2}\right)-k_{z}v\tan\left(k\frac{\Delta t}{2}\right)\cos\left(\omega\frac{\Delta t}{2}\right)\right)/k^{2}k_{x}\Delta t,\label{eq:C3x}
\end{multline}
\begin{multline}
C_{2z}=-k_{z}^{\prime}\sum_{m_{x}}\left(S^{J}\right)^{2}\\
\left(\zeta_{z}\sin\left(k\,\Delta t\right)k_{x}^{2}\sin\left(\omega\frac{\Delta t}{2}\right)-k\, k_{z}^{2}\Delta t\, C_{0}\right)/k^{3}k_{z}\Delta t,\label{eq:C2z}
\end{multline}
\begin{equation}
C_{3z}=k_{z}^{\prime}\Delta t^{2}\sum_{m_{x}}\left(S^{J}\right)^{2}\left(\zeta_{z}k_{x}^{2}+\zeta_{x}k_{z}^{2}\right)/4k^{2}k_{z}.\label{eq:C3z}
\end{equation}
For $\gamma^{2}$ large but not infinite, which is the focus of this
paper, the approximate solutions of Eq. \ref{eq:drformfull} are the
solutions of
\begin{equation}
C_{0}+n\sum_{m_{z}}C_{1}\csc\left[\left(\omega-k_{z}^{\prime}v\right)\frac{\Delta t}{2}\right]+n\sum_{m_{z}}C_{2x}\csc^{2}\left[\left(\omega-k_{z}^{\prime}v\right)\frac{\Delta t}{2}\right]=0,\label{eq:drform}
\end{equation}
plus an additional, stable mode,
\begin{equation}
\omega=k_{z}^{\prime}v-\left.\frac{2}{\Delta t}\gamma^{-2}n\, C_{3z}C_{3x}/C_{2x}\right|_{\omega=k_{z}^{\prime}v},\label{eq:extramode}
\end{equation}
provided that $C_{2x}$ does not vanish there. If it does, the extra
mode may be unstable, with growth rate scaling as $\gamma^{-1}$.
As already noted, sums over $m_{x}$ can be performed explicitly,
\begin{multline}
\sum_{m_{x}}\left(S^{J}\right)^{2}=-\left[\sin\left(k_{z}^{\prime}\frac{\Delta z}{2}\right)/\left(k_{z}^{\prime}\frac{\Delta z}{2}\right)\right]^{2\ell_{z}+2}\\
\frac{1}{\left(2\ell_{x}+1\right)!}\left[\sin\left(k_{x}\frac{\Delta x}{2}\right)\right]^{2\ell_{x}+2}\left.\frac{d^{2\ell_{x}+1}\cot\left(\kappa\right)}{d\,\kappa^{2\ell_{x}+1}}\right|_{\kappa=k_{x}\frac{\Delta x}{2}},\label{eq:summx}
\end{multline}
 
\begin{multline}
\sum_{m_{x}}k_{x}^{\prime}\left(S^{J}\right)^{2}=\left[\sin\left(k_{z}^{\prime}\frac{\Delta z}{2}\right)/\left(k_{z}^{\prime}\frac{\Delta z}{2}\right)\right]^{2\ell_{z}+2}\\
\frac{1}{\left(2\ell_{x}\right)!}\left[\sin\left(k_{x}\frac{\Delta x}{2}\right)\right]^{2\ell_{x}+2}\left.\frac{d^{2\ell_{x}}\cot\left(\kappa\right)}{d\,\kappa^{2\ell_{x}}}\right|_{\kappa=k_{x}\frac{\Delta x}{2}}.\label{eq:summx-kxp}
\end{multline}
Analogous expressions for the $C$'s also can be obtained for the
PSATD-conventional algorithm. 

Vacuum electromagnetic modes are described by $C_{0}=0,$
\begin{equation}
\sin^{2}\left(\omega\frac{\Delta t}{2}\right)=\sin^{2}\left(k\frac{\Delta t}{2}\right),\label{eq:vacuumDR}
\end{equation}
which yields real $\omega$ for all values of $k\,\Delta t$. The
PSATD algorithm thus has no Courant limit on $\Delta t$. However,
$\left|\omega\right|$ begins decreasing with increasing $k\,\Delta t$,
when $k\,\Delta t$ first exceeds $2\pi$. This threshold is expressed
in terms of the grid cell size as $\Delta t>\Delta t_{c}=\left(\Delta z^{-2}+\Delta x^{-2}\right)^{-\nicefrac{1}{2}}$,
which is recognizable as the usual Courant condition in FDTD algorithms.
Digitally filtering wave-numbers for which $k>2\pi/\Delta t_{c}$
often is prudent.

\begin{figure}
\centering{}\includegraphics[bb=0bp 0bp 360bp 0bp,scale=0.93]{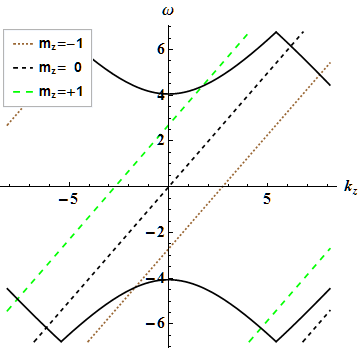}\caption{\label{fig:Normal-mode-diagram}PSATD normal mode diagram for $v\triangle t/\triangle z=1.2$
and $k_{x}=\pi/2\triangle x$, showing electromagnetic modes (numerically
distorted for $k>\pi/\triangle t$ ) and spurious beam modes, $m_{z}=\left[-1,\,1\right]$.
Numerical Cherenkov instabilities are strongest near mode intersections.}
\end{figure}

\begin{figure}
\centering{}\includegraphics[scale=0.93]{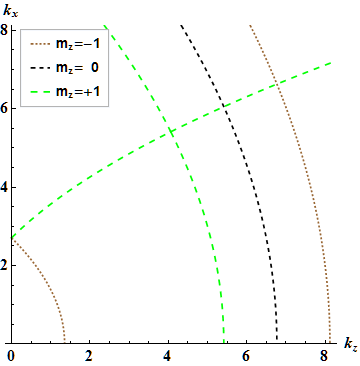}\caption{\label{fig:Resonance curves}Locations in \textit{k}-space of PSATD
resonances between electromagnetic modes and spurious beam modes,
$m_{z}=\left[-1,\,+1\right]$, for $v\triangle t/\triangle z=1.2$.
Intersecting resonance curves occur at different frequencies and,
therefore, do not interact.}
\end{figure}

All beam modes in Eq.(\ref{eq:drformfull}) are numerical artifacts,
even the $m_{z}=0$ mode, and their interaction with the electromagnetic
modes gives rise to the numerical Cherenkov instability \citep{godfrey1974numerical,godfrey1979electro}.
Fig. \ref{fig:Normal-mode-diagram} is a typical normal mode diagram,
showing the two electromagnetic modes and beam aliases $m_{z}=[-1,\,+1]$
for $v\,\Delta t/\Delta z=1.2$ and $k_{x}=\nicefrac{1}{2}\pi/\triangle x$.
(Unless otherwise noted, other parameters for these and other figures
are $n=1$ and $\Delta x=\Delta z=0.3868$.) Not surprisingly, most
rapid growth occurs at resonances, where normal modes intersect. Fig.
\ref{fig:Resonance curves} depicts the locations in \textit{k}-space
of normal mode intersections, such as those in Fig. \ref{fig:Normal-mode-diagram},
as $k_{x}$ is varied.%
\footnote{Software to generate plots such as those in Figs. \ref{fig:Normal-mode-diagram}
and \ref{fig:Resonance curves} is available in Computable Document
Format \citep{WolframCDF} at http://hifweb.lbl.gov/public/BLAST/Godfrey/.%
} Because the electromagnetic modes are dispersionless for $\Delta t<\Delta t_{c}$,
the otherwise often dominant $m_{z}=0$ numerical Cherenkov instability
cannot occur unless $\Delta t$ somewhat exceeds $\Delta t_{c}$.
To be precise, the resonant $m_{z}=0$ instability occurs only for
$\nicefrac{\Delta t}{\Delta x}>2\left(\nicefrac{\Delta x}{\Delta t_{c}}-\nicefrac{\Delta z}{\Delta x}\right)$,
or $\nicefrac{\Delta t}{\Delta z}>2\left(\sqrt{2}-1\right)$ for $\Delta x=\Delta z$
(accurate to order $\gamma^{-2}$). The $m_{z}=-1$ instability dominates
at smaller time-steps.

More generally, instability resonances occur at 
\begin{equation}
k_{x}^{r}=\left(\left(\left(k_{z}+m_{z}\frac{2\pi}{\triangle z}\right)v-p\frac{2\pi}{\triangle t}\right)^{2}-k_{z}^{2}\right)^{\nicefrac{1}{2}},\label{eq:krres}
\end{equation}
where \emph{p} is any integer within the domain,
\begin{equation}
\left[m_{z}v\frac{\triangle t}{\triangle z}-\frac{\triangle t}{2\triangle x}\:,\:\left(m_{z}+\frac{1}{2}\right)v\frac{\triangle t}{\triangle z}+\left(\triangle z^{-2}+\triangle x^{-2}\right)^{\nicefrac{1}{2}}\frac{\triangle t}{2}\right],\label{eq:p}
\end{equation}
except $p=0$ for $m_{z}=0$. In effect, \emph{p} is the temporal
alias number.

Ref. \citep{godfrey2013esirkepov} described in detail how to estimate
numerical Cherenkov instability peak growth rates as a function of
$\Delta t$, based on an approximate evaluation of Eq. \ref{eq:drformfull}.
This approach was used to good effect to explain the existence and
value of time-steps for which instability growth rates in WARP-FDTD
\citep{VayJCP2011,Vay2012AAC} and other PIC codes (\emph{e.g.}, \citep{Xu2012,UCLA2012AAC})
were greatly reduced. Here, the approximate resonant growth rate is
\begin{equation}
Im\left(\omega\right)\simeq\left|n\, C_{2}\Delta t/4k_{z}\right|^{\nicefrac{1}{3}}/\Delta t,\label{eq:resonant}
\end{equation}
with $C_{2}$ evaluated at $\omega=k_{z}v$ and $k_{x}$ chosen to
satisfy the resonance condition.

\begin{figure}
\centering{}\includegraphics[scale=0.93]{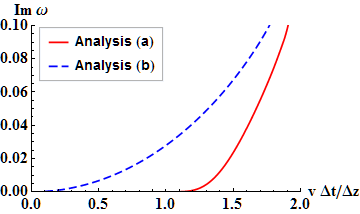}\caption{\label{fig:Stability criterion}Approximate maximum growth rates for
PSATD options (a) and (b) with cubic interpolation and digital filtering.
Option (c) exhibits zero growth in this approximation.}
\end{figure}

In this paper we focus instead on finding parameters for which non-resonant
growth at small $k$ naturally is minimized, while relying on digital
filtering to suppress the otherwise faster growing resonant instabilities
at large $k$. Non-resonant instability occurs when $C_{0}C_{2}>nC_{1}^{\,2}/4$,
evaluated at $\omega\simeq k_{z}^{\prime}$ and arbitrary $k_{x}$.
The resulting growth rate is
\begin{equation}
Im\left(\omega\right)\simeq\frac{\sqrt{4nC_{0}C_{2}-n^{2}C_{1}^{\,2}}}{C_{0}\Delta t}.\label{eq:quadratic growth}
\end{equation}
By numerical experimentation we have found that $C_{0}C_{2}>0$ is
bounded away from $k_{z}=0$ for $m_{z}=0$ and any $\zeta_{z}<1$,
and that for some choices of $\zeta_{z}$ this region free of non-resonant
instability can be fairly large. Fig. \ref{fig:Stability criterion}
depicts maximum approximate growth rates as a function of $v\,\Delta t/\Delta z$
according to Eq. \ref{eq:quadratic growth} for the PSATD-Esirkepov
algorithm with (a) $\zeta_{z}=\left(k_{z}\triangle z/2\right)\cot\left(k_{z}\triangle z/2\right),\zeta_{x}=\left(k_{x}\triangle x/2\right)\cot\left(k_{x}\triangle x/2\right)$
or (b) $\zeta_{z}=\zeta_{x}=1$, cubic interpolation, and smoothing
as in Eq. (37) of \citep{godfrey2013esirkepov}. Option (a) exhibits
essentially no instability for $v\,\Delta t/\Delta z<1.2$, while
option (b) does exhibit instability there. This useful finding is
substantiated in Sec. 6. Incidentally, both numerical and analytical
solutions of Eq. (\ref{eq:drformfull}) indicate significant numerical
instability even at small $k_{z}$ when $\zeta_{x}>1$.

Of course, other fruitful choices for $\zeta_{z}$ may exist. One
promising possibility is $\zeta_{z}$ chosen such that $C_{2}$ vanishes
for $\omega=k_{z}v$, in order to suppress non-resonant $m_{z}=0$
growth in accordance with Eq. (\ref{eq:quadratic growth}),
\begin{multline}
\zeta_{z}=k\, k_{z}\triangle t\left(\sin^{2}\left(k_{z}\frac{\Delta t}{2}\right)-\sin^{2}\left(k\frac{\Delta t}{2}\right)\right)\csc\left(k_{z}\frac{\Delta t}{2}\right)\csc\left(k\frac{\Delta t}{2}\right)/\\
2\left(k_{z}\sin\left(k_{z}\frac{\Delta t}{2}\right)\cos\left(k\frac{\Delta t}{2}\right)-k\,\cos\left(k_{z}\frac{\Delta t}{2}\right)\sin\left(k\frac{\Delta t}{2}\right)\right)\label{eq:resz}
\end{multline}
Equivalently, Eq. (\ref{eq:resz}) is obtained by setting to zero
the first term in the Laurent expansion of Eq. (\ref{eq:drform})
about $\omega=k_{z}v$. Note that $v_{z}$ has been set equal to unity
in this expression to assure that $\zeta_{z}\rightarrow1$ as $k\rightarrow0$.
Moreover, it is necessary to impose $0\leq\zeta_{z}\leq1$. We do
this by setting $\zeta_{z}=0$ everywhere that the constraint just
given is not satisfied, which is almost everywhere outside the curve,
$k_{z}=\frac{\pi}{\Delta t}-k_{x}^{2}\frac{\Delta t}{4\pi}$. Not
coincidentally, this is the curve at which the first $m_{z}=0$ instability
resonance occurs. Seemingly, the corresponding $\zeta_{x}$ should
be obtained by setting to zero the second term in the Laurent expansion,
\begin{multline}
\zeta_{x}=k\, k_{x}^{2}\triangle t\left(k\,\sin\left(k_{z}\frac{\Delta t}{2}\right)\sin\left(k\frac{\Delta t}{2}\right)\left(\cos^{2}\left(k_{z}\frac{\Delta t}{2}\right)+\cos^{2}\left(k\frac{\Delta t}{2}\right)\right)\right.\\
-\left.k_{z}\cos\left(k_{z}\frac{\Delta t}{2}\right)\cos\left(k\frac{\Delta t}{2}\right)\left(\sin^{2}\left(k_{z}\frac{\Delta t}{2}\right)+\sin^{2}\left(k\frac{\Delta t}{2}\right)\right)\right)\csc\left(k\frac{\Delta t}{2}\right)/\\
2k_{z}\cos\left(k_{z}\frac{\Delta t}{2}\right)\left(k_{z}\sin\left(k_{z}\frac{\Delta t}{2}\right)\cos\left(k\frac{\Delta t}{2}\right)-k\,\cos\left(k_{z}\frac{\Delta t}{2}\right)\sin\left(k\frac{\Delta t}{2}\right)\right)^{2}.\label{eq:resx}
\end{multline}
However, it satisfies the constraint, $0\leq\zeta_{x}\leq1$, over
too small a region in \emph{k}-space. Credible alternatives are $\zeta_{x}=1$,
$\zeta_{x}=\left(k_{x}\triangle x/2\right)\cot\left(k_{x}\triangle x/2\right)$,
and $\zeta_{x}=\zeta_{z}$. Each produces roughly the same growth
rates when paired with Eq. (\ref{eq:resz}), at least when digital
filtering is employed as well. The choice, $\zeta_{x}=\zeta_{z}$,
is designated PSATD option (c) and used in representative numerical
calculations in Sec. 6. Although this approach may seem rather arbitrary,
it does give good results.

Axial group velocities of unstable modes, $v_{g}=\partial\omega/\partial k_{z}$,
are of interest when dealing with short beam pulses, because numerical
instability energy propagates backward relative to the beam pulse,
limiting total growth, when the instability group velocity is somewhat
less than the beam velocity. Low instability group velocities can
be expected for beam aliases interacting resonantly with backward
propagating electromagnetic waves. Indeed, numerical solutions to
the dispersion relation predict group velocities between 0.3 and 0.5
the beam velocity in this case. On the other hand, numerical instabilities
associated with beam aliases interacting with forward propagating
electromagnetic waves can be expected to have group velocities about
equal to the beam velocity. The same is true of non-resonant instabilities,
and numerical solutions of the dispersion relation corroborate these
expectations.

\begin{figure}
\centering{}\includegraphics[scale=0.93]{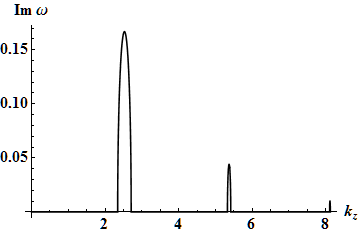}\caption{\label{fig:grow1d}PSATD one-dimensional growth rate for $m_{z}=0$
and $v\,\triangle t/\triangle z=3$.}
\end{figure}

The PSATD-Esirkepovk one-dimensional dispersion relation is obtained
simply by setting $k_{x}$ to zero in Eqs. (\ref{eq:C0}) - (\ref{eq:C2x}),
yielding 
\begin{equation}
C_{0}=\left[\omega\right]^{2}-\left[k_{z}\right]^{2},\label{eq:C0-1d}
\end{equation}
\begin{multline}
C_{1}=-\zeta_{x}\left(S^{J}\right)^{2}\cos\left(k_{z}^{\prime}v\frac{\Delta t}{2}\right)\sin\left(k_{z}\Delta t\right)\\
\left(\sin\left(\omega\frac{\Delta t}{2}\right)-v\tan\left(k_{z}\frac{\Delta t}{2}\right)\cos\left(\omega\frac{\Delta t}{2}\right)\right)/k_{z}\Delta t,\label{eq:C1-1d}
\end{multline}
and $C_{2}=0$. Resonant instability occurs when $-n\, C_{1}/\sin\left(\omega\Delta t\right)\cos\left[\left(\omega-k_{z}^{\prime}v\right)\frac{\Delta t}{2}\right]>0$,
evaluated at the resonance frequency, in which case the growth rate
is the square root of that quantity. Interestingly, for $m_{z}=0$
in the limit $v\rightarrow1$, the Cherenkov resonance drops out,
and the dispersion relation simplifies further to
\begin{equation}
\sin^{2}\left(\omega\frac{\Delta t}{2}\right)-\sin^{2}\left(k_{z}\frac{\Delta t}{2}\right)-n\zeta_{x}\Delta t\left(S^{J}\right)^{2}\sin\left(k_{z}\Delta t\right)/4k_{z}=0.\label{eq:dr-v11d}
\end{equation}
Nonetheless, an instability still occurs approximately where the resonance
would have been, namely $k_{z}$ just less than an integer multiple
of $\pi/\Delta t$. The numerical solution of Eq. (\ref{eq:dr-v11d})
for $v\,\Delta t/\Delta z=3$ is provided in Fig. \ref{fig:grow1d}.
Peak growth for $k_{x}=0$ in this case is only about one-third the
peak growth at finite $k_{x}$.

\section{Numerical solutions}

\begin{figure}
\centering{}\includegraphics[scale=0.75]{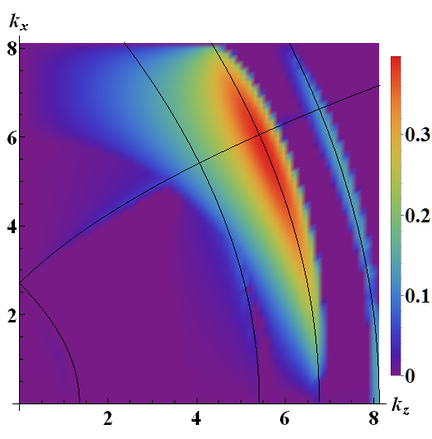}\caption{\label{fig:growContour}Growth rates from PSATD dispersion relation
for option (a), $m_{z}=\left[-1,\,+1\right]$, and $v\triangle t/\triangle z=1.2$.
Superimposed are the resonance curves from Fig. \ref{fig:Resonance curves}}
\end{figure}

Numerical solutions to the complete linear dispersion relations, presented
in Sec. 4, and instability growth rates measurements from corresponding
WARP simulations were performed as described in Sec. 5 of \citep{godfrey2013esirkepov}.
A typical dispersion relation growth spectrum, in this case corresponding
to the parameters of Fig. \ref{fig:Resonance curves} with option
(a), $\zeta_{z}=\left(k_{z}\triangle t/2\right)\cot\left(k_{z}\triangle t/2\right)$
and $\zeta_{x}=\left(k_{x}\triangle t/2\right)\cot\left(k_{x}\triangle t/2\right)$,
is depicted in Fig. \ref{fig:growContour}. Growth is dominated by
the $m_{z}=0$ numerical instability. Note that non-resonant growth
associated with the $m_{z}=0$ mode is bounded well away from small
$k_{z}$, as predicted in the previous Section. The instability group
velocity is about 0.5 on resonance and 1.0 well off resonance.

\begin{figure}
\centering{}\includegraphics[scale=0.95]{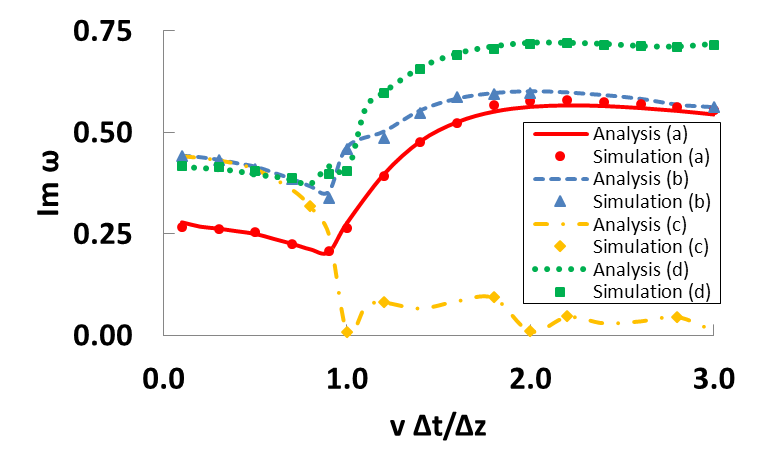}\caption{\label{fig:linear-nosmooth}Maximum growth rates for PSATD options
(a), (b), (c), and (d) with linear interpolation and no digital filtering.
Markers represent corresponding simulation results.}
\end{figure}

Fig. \ref{fig:linear-nosmooth} plots maximum growth rates versus
$v\,\Delta t/\Delta z$ for options (a), (b), and (c), as well as
for option (d), which is PSATD with conventional current interpolation.
Recall that the option (d) dispersion relation is given by Eqs. (\ref{eq:M3x3})
and (\ref{eq:Mzz-alt}) - (\ref{eq:Mxy-alt}). (A summary of the options
is given in Table \ref{Tableoption}.) Growth rates for option (a)
are noticeably smaller than those for options (b) and (d) with $v\,\Delta t/\Delta z$
less than about 1.5, in part because $\zeta_{z}$ introduces smoothing
at large $k$, which is where the dominant resonances occur in this
range of time-steps. On the other hand, the curves for options (a)
and (b) converge for large $v\,\Delta t/\Delta z$, because $\zeta_{z}$
for both options (and indeed for all valid choices of $\zeta_{z}$)
approaches unity at small $k$, which is where the dominant resonances
occur at large time-steps. An inflection occurs in curves (a), (b),
and (d) near $v\,\Delta t/\Delta z\approx0.9$, where the $m_{z}=0$
resonant instability begins to dominate the $m_{z}=-1$ and other
resonances. PSATD option (c), designed to suppress the the $m_{z}=0$
instability, both resonant and non-resonant, is seen to do so quite
effectively. Growth plummets to near zero at $v\,\Delta t/\Delta z=1$
and is modestly larger at larger values of $v\,\Delta t/\Delta z$
due only to residual $m_{z}=\pm1$ resonant instabilities. Agreement
between theory and simulation growth rates is very good in all cases.
The simulation growth rate measurements themselves appear to be accurate
to better than 2\%, except perhaps for very small growth rates.

\begin{table}

\caption{Algorithm options used in Fig. \ref{fig:linear-nosmooth}, \ref{fig:linear-filtering},
\ref{fig:cubic-filtering}, and elsewhere.}

\centering{}\label{Tableoption}%
\begin{tabular}{|c|c|c|}
\hline 
Option & Current Factors or Equations & Comments\tabularnewline
\hline 
\hline 
(a) & %
\begin{tabular}{c}
$\zeta_{x}=\left(k_{x}\triangle x/2\right)\cot\left(k_{x}\triangle x/2\right)$\tabularnewline
$\zeta_{z}=\left(k_{z}\triangle z/2\right)\cot\left(k_{z}\triangle z/2\right)$\tabularnewline
\end{tabular} & Equivalent to Esirkepov in real space\tabularnewline
\hline 
(b) & $\zeta_{z}=\zeta_{x}=1$ & Esirkepov in \emph{k}-space (base case)\tabularnewline
\hline 
(c) & $\zeta_{x}=\zeta_{z}$, as defined in Eq. (\ref{eq:resz}) & Reduces order of nonphysical resonances\tabularnewline
\hline 
(d) & Eqs. (\ref{eq:M3x3}), (\ref{eq:Mzz-alt}) - (\ref{eq:Mxy-alt}) & Conventional current deposition at nodes\tabularnewline
\hline 
\end{tabular}
\end{table}

\begin{figure}
\centering{}\includegraphics[scale=0.45]{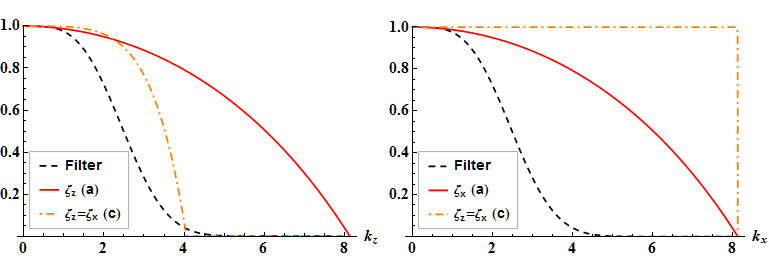}\caption{\label{fig:smooth}Left: $k_{z}$-dependent factor of ten-pass bilinear
filter, $\zeta_{z}$ for option (a) (which depends only on $k_{z}$),
and $\zeta_{z}=\zeta_{x}$ for option (c) evaluated at $k_{x}=0$
and $\triangle t/\triangle z=2$. Right: \emph{$k_{x}$}-dependent
factor of ten-pass bilinear filter, $\zeta_{x}$ for option (a) (which
depends only on \emph{$k_{x}$}), and $\zeta_{z}=\zeta_{x}$ for option
(c) evaluated at $k_{z}=0$ and $\triangle t/\triangle z=2$.}
\end{figure}

\begin{figure}
\centering{}\includegraphics[scale=0.95]{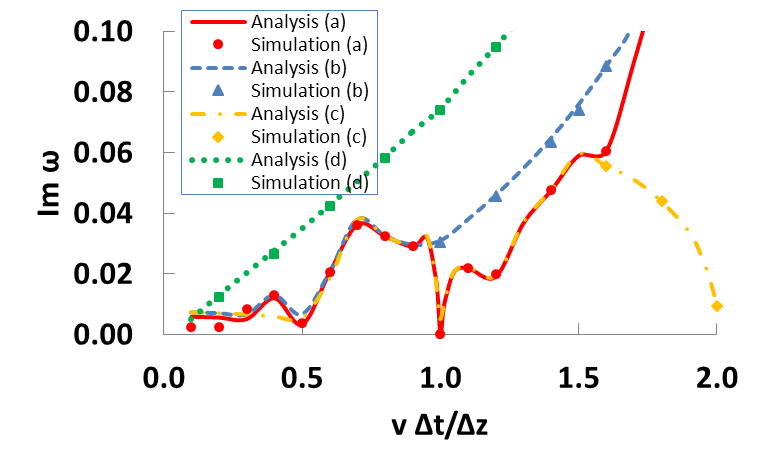}\caption{\label{fig:linear-filtering}Maximum growth rates for PSATD options
(a), (b), (c), and (d) with linear interpolation and digital filtering.
Markers represent corresponding simulation results.}
\end{figure}

As explained in the previous Section, PSATD combined with digital
filtering can be very effective at suppressing the numerical Cherenkov
instability. Since filtering can be applied directly in \emph{k}-space,
any suitable filtering profile can be employed in a straightforward
manner. (Digital filtering of the numerical Cherenkov instabiity in
FDTD algorithms is described in \citep{godfrey2013esirkepov,GreenwoodJCP04}.)
To facilitate comparison with earlier analysis for WARP-FDTD \citep{godfrey2013esirkepov},
we use the same ten-pass (including two compensation passes) bilinear
filter used there. The \emph{$k_{z}$-} and $k_{x}$-dependent factors
of the filter function are displayed in Fig. \ref{fig:smooth}. (Also
shown are $\zeta_{z}$ and $\zeta_{x}$ for options (a) and (c). Remember,
however, that these current multipliers are not equivalent to digital
filters, although they can introduce a degree of smoothing.) Applying
this filter with parameters otherwise identical to those in Fig. \ref{fig:linear-nosmooth}
reduces growth rates by a factor of five or so over the range of $v\,\Delta t/\Delta z$
shown in Fig. \ref{fig:linear-filtering} (or for $v\,\Delta t/\Delta z$<1
in the case of option (c), which has small growth for larger time-steps
even without filtering). At larger time-steps growth rates increase
toward their unfiltered values, as the dominant resonant modes move
to progressively smaller $k_{z}$. (For instance, the option (a) filtered
maximum growth rate increases to 74\% of its unfiltered value by $v\,\Delta t/\Delta z=3$.)
Maximum growth rates oscillate irregularly for options (a) and (c)
when $v\,\Delta t/\Delta z$ is less than about 1.3, and for option
(b) when it is less than about 1.0, as higher order resonances move
through the weakly filtered region at small $k$. Digital filtering
seems less effective for option (d), probably because its $m_{z}=0$
non-resonant growth at small \emph{k} is larger than in the other
options.

\begin{figure}
\centering{}\includegraphics[scale=0.95]{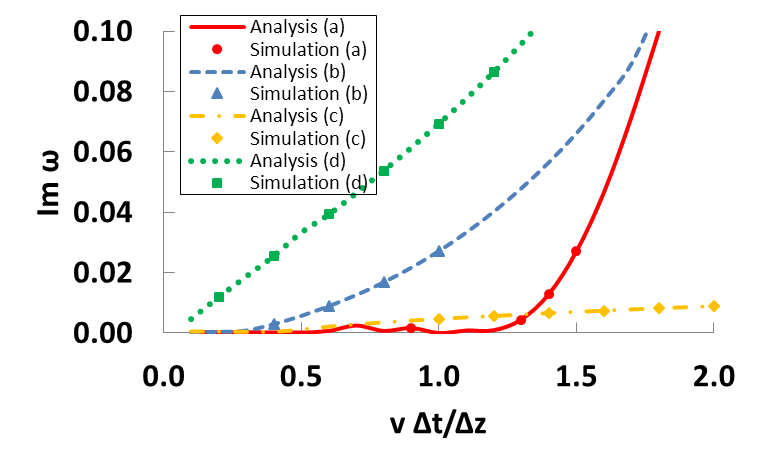}\caption{\label{fig:cubic-filtering}Maximum growth rates for PSATD options
(a), (b), (c), and (d) with cubic interpolation and digital filtering.
Markers represent corresponding simulation results.}
\end{figure}

The weak instability growth for options (a) and (c) can be further
reduced by higher order interpolation. As illustrated in Fig. \ref{fig:cubic-filtering}
and, with slightly less accuracy, in Fig. \ref{fig:Stability criterion},
cubic interpolation almost completely eliminates numerical Cherenkov
instability growth in option (a) for $v\,\Delta t/\Delta z<1.3$.
Option (c) performs almost as well in that same time-step range and
much better outside it. Quadratic interpolation performs almost as
well as cubic in this regard. Incidentally, the residual instability
for option (c) is a finite $\gamma$ effect, dropping to zero for
infinite $\gamma$.

One might reasonably ask whether the superior stability properties
of option (c) at larger time steps are due only to the digital filtering
of the transverse currents that it entails. No, is the answer, as
can be demonstrated from numerical solution of the option (b) dispersion
relation with the right side of Eq. (\ref{eq:resz}) used as a digital
filter applied to \emph{n} throughout. Doing so effectively suppresses
the $m_{z}=0$ resonant instability but not its non-resonant counterpart,
with maximum growth rates at larger time steps of order one-third
those of option (b) without digital filtering, Fig. \ref{fig:linear-nosmooth}.
And, when digital filtering equal to the right side of Eq. (\ref{eq:resz})
is combined with the digital filtering already employed in Fig. \ref{fig:linear-filtering}
or \ref{fig:cubic-filtering}, the results are practically indistinguishable
from those of option (b).

\begin{figure}
\centering{}\includegraphics[scale=0.95]{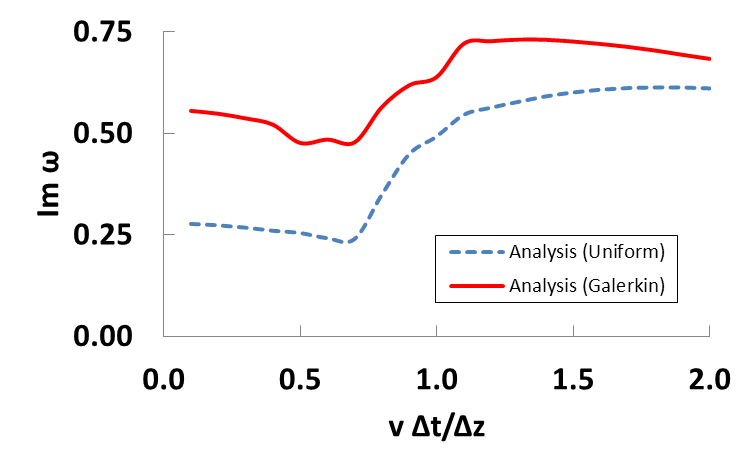}\caption{\label{fig:uniform-galerkin}Maximum growth rates for PSATD Uniform
and Galerkin linear interpolation schemes and no digital filtering.}
\end{figure}

The PSATD algorithm also accommodates field interpolation using the
Galerkin and Uniform schemes discussed in \citep{godfrey2013esirkepov}.
(Fields are computed at mesh points as described in Sec. 2 and then
averaged to the staggered Yee mesh\citep{Yee}.) Results for these
two schemes with linear interpolation and no digital filtering are
provided in Fig. \ref{fig:uniform-galerkin}. Both exhibit non-resonant
instability growth rates at small \emph{k}. Consequently, there appears
to be no advantage in using these more complicated field interpolation
approaches with PSATD.

\begin{figure}
\begin{centering}
\includegraphics[scale=0.95]{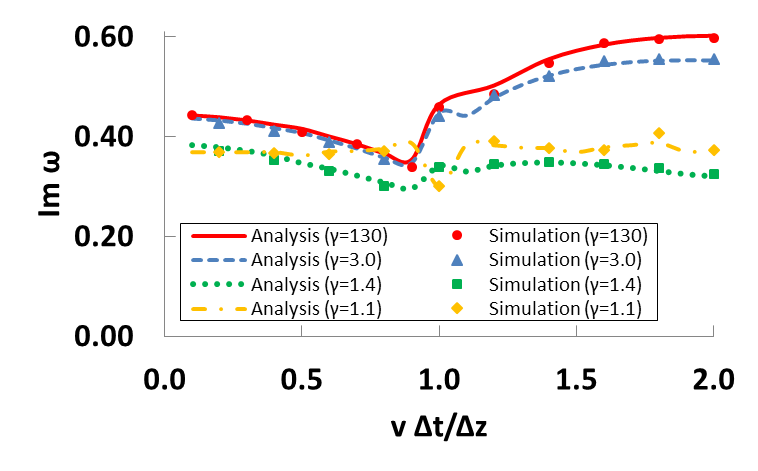}
\par\end{centering}

\caption{\label{fig:Low gamma}Maximum growth rates for PSATD option (b) with
$\gamma=130,\,3.0,\,1.4,\,1.1$, linear interpolation, and no filtering.
Markers represent corresponding simulation results.}
\end{figure}

Although Figs. \ref{fig:linear-nosmooth}, \ref{fig:linear-filtering},
and \ref{fig:cubic-filtering} demonstrate clearly the validity of
the numerical dispersion relation in the large $\gamma$ limit, they
indicate little about its validity more generally. We have, therefore,
run comparisons between the dispersion relation and WARP-PSATD option
(b) simulations for $\gamma=3.0,\,1.4,\,1.1$ with linear interpolation
and no digital filtering. Once again, agreement is excellent; see
Fig. (\ref{fig:Low gamma}). Maximum growth rates for $\gamma$ as
low as 3 are essentially the same as those for $\gamma=130$. However,
the k-space spectrum at $\gamma=3.0$ also shows signs of the well
known $m_{z}=-1$ quasi-one-dimensional, electrostatic numerical instability
\citep{Langdon1970,Okuda1970}. For smaller $\gamma$ yet, the numerical
Cherenkov instability growth rate decreases modestly, while the electrostatic
numerical instability growth rate increases as $1/\gamma$ for fixed
\emph{n}. (Recall that n is defined in this paper as the density divided
by $\gamma$.) The two become comparable at $\gamma\approx1.4$, and
the electrostatic instability dominates strongly at $\gamma=1.1$.
The electrostatic numerical instability can be suppressed by using
any field interpolation algorithm that offsets $E_{z}$ by $\triangle z/2$
relative to $\varrho$ (or $\mathbf{W}$ in the Esirkepov current
algorithm) and interpolates it with a spline one order lower in z
relative to $\varrho$ or to $\mathbf{W}$ \citep{lewis1972variational,langdon1973energy},
such as the Galerkin ``energy-conserving'' algorithm. Even the Uniform
algorithm ameliorates the electrostatic instability to a degree, stabilizing
the strong $m_{z}=-1$ mode but destabilizing the slower $m_{z}=+1$
mode. Of course, digital filtering plus cubic interpolation also works
well.

\section{PSTD stability results}

The numerical stability properties of the related PSTD algorithm \citep{Liu1997}
recently were addressed in \citep{Xu2012}. Here, we focus on comparison
of PSATD and PSTD growth rates. The PSTD dispersion relation can be
derived following the procedures used to analyze the PSATD algorithm.
Under the assumptions leading to Eqs. (\ref{eq:leapfrog-alt1}), (\ref{eq:leapfrogB}),
and (\ref{eq:Bave}), the corresponding PSTD equations are 
\begin{multline}
\mathbf{E}^{n+1}=\mathbf{E}^{n}-i\mathbf{k}\times\mathbf{B}^{n+\nicefrac{1}{2}}\triangle t-\mathbf{\mathbf{\zeta}:}\mathbf{J}_{e}^{n+\nicefrac{1}{2}}\triangle t\\
+\mathbf{k}\mathbf{k}\cdot\mathbf{\mathbf{\zeta}:}\mathbf{J}_{e}^{n+\nicefrac{1}{2}}\triangle t/k^{2}-\mathbf{k}\mathbf{k}\cdot\mathbf{J}_{e}^{n+\nicefrac{1}{2}}\triangle t/k^{2},\label{eq:EleapfrogPSTD}
\end{multline}
\begin{equation}
\mathbf{B}^{n+\nicefrac{3}{2}}=\mathbf{B}^{n+\nicefrac{1}{2}}+i\mathbf{k}\times\mathbf{E}^{n+1}\triangle t,\label{eq:BleapfrogPSTD}
\end{equation}
\begin{equation}
\mathbf{B}^{n}=\left(\mathbf{B}^{n+\nicefrac{1}{2}}+\mathbf{B}^{n-\nicefrac{1}{2}}\right)/2.\label{eq:BavePSTD}
\end{equation}
 As noted in \citep{Vay2013PSATD}, these equations also can be obtained
by expanding their PSATD counterparts to first order in \emph{k}.

The dispersion relation again takes the form of (\ref{eq:M3x3}),
but with $[\mathbf{k}]=\mathbf{k}$,
\begin{equation}
\xi_{z,z}=-n\gamma^{-2}\sum_{m}S^{J}S^{E_{z}}\csc^{2}\left[\left(\omega-k_{z}^{\prime}v\right)\frac{\Delta t}{2}\right]\left(k_{z}^{2}+\zeta_{z}k_{x}^{2}\right)\Delta t^{2}\left[\omega\right]k_{z}^{\prime}/4k^{2}k_{z}\mathrm{,}\label{eqMzzPSTD}
\end{equation}
\begin{equation}
\xi_{z,x}=-n\sum_{m}S^{J}S^{E_{x}}\csc\left[\left(\omega-k_{z}^{\prime}v\right)\frac{\Delta t}{2}\right]\eta_{z}\Delta t\, k_{x}^{\prime}/2k^{2}k_{z},\label{eq:MzxPSTD}
\end{equation}
\begin{equation}
\xi_{z,y}=nv\sum_{m}S^{J}S^{B_{y}}\csc\left[\left(\omega-k_{z}^{\prime}v\right)\frac{\Delta t}{2}\right]\eta_{z}\Delta t\, k_{x}^{\prime}/2k^{2}k_{z},\label{eq:MzyPSTD}
\end{equation}
\begin{equation}
\xi_{x,z}=-n\gamma^{-2}\sum_{m}S^{J}S^{E_{z}}\csc^{2}\left[\left(\omega-k_{z}^{\prime}v\right)\frac{\Delta t}{2}\right]\left(1-\zeta_{z}\right)\Delta t^{2}\left[\omega\right]k_{x}k_{z}^{\prime}/4k^{2},\label{eq:MxzPSTD}
\end{equation}
\begin{equation}
\xi_{x,x}=-n\sum_{m}S^{J}S^{E_{x}}\csc\left[\left(\omega-k_{z}^{\prime}v\right)\frac{\Delta t}{2}\right]\eta_{x}\Delta t\, k_{x}^{\prime}/2k^{2}k_{x},\label{eq:MxxPSTD}
\end{equation}
\begin{equation}
\xi_{x,y}=nv\sum_{m}S^{J}S^{B_{y}}\csc\left[\left(\omega-k_{z}^{\prime}v\right)\frac{\Delta t}{2}\right]\eta_{x}\Delta t\, k_{x}^{\prime}/2k^{2}k_{x},\label{eq:MxyPSTD}
\end{equation}
and
\begin{equation}
\eta_{z}=\cot\left[\left(\omega-k_{z}^{\prime}v\right)\frac{\Delta t}{2}\right]\left(k_{z}^{2}+\zeta_{z}k_{x}^{2}\right)\sin\left(k_{z}^{\prime}v\frac{\Delta t}{2}\right)+\left(1-\zeta_{x}\right)k_{z}^{2}\cos\left(k_{z}^{\prime}v\frac{\Delta t}{2}\right),\label{eq:etazPSTD}
\end{equation}
\begin{equation}
\eta_{x}=\cot\left[\left(\omega-k_{z}^{\prime}v\right)\frac{\Delta t}{2}\right]\left(1-\zeta_{z}\right)k_{x}^{2}\sin\left(k_{z}^{\prime}v\frac{\Delta t}{2}\right)+\left(k_{x}^{2}+\zeta_{x}k_{z}^{2}\right)\cos\left(k_{z}^{\prime}v\frac{\Delta t}{2}\right).\label{eq:etaxPSTD}
\end{equation}

Provided that currents and fields are interpolated to or from the
same mesh points, the high-$\gamma$ dispersion relation again takes
the form in Eq. \ref{eq:drformfull} with $C_{0}$ as before and
\begin{multline}
C_{1}=-\sum_{m_{x}}k_{x}^{\prime}\left(S^{J}\right)^{2}\cos\left(k_{z}^{\prime}v\frac{\Delta t}{2}\right)\\
\left\{ 2\zeta_{x}k_{z}\left(2k_{z}\sin\left(\omega\frac{\Delta t}{2}\right)-k^{2}v\,\Delta t\cos\left(\omega\frac{\Delta t}{2}\right)\right)+k_{x}^{2}\Delta t^{2}C_{0}/\sin\left(\omega\frac{\Delta t}{2}\right)\right\} /4k^{2}k_{x},\label{eq:C1PSTD}
\end{multline}
\begin{multline}
C_{2}=k_{x}\sum_{m_{x}}k_{x}^{\prime}\left(S^{J}\right)^{2}\cos\left[\left(\omega-k_{z}^{\prime}v\right)\frac{\Delta t}{2}\right]\sin\left(k_{z}^{\prime}v\frac{\Delta t}{2}\right)\\
\left\{ 2\zeta_{z}k_{x}\left(2k_{z}\sin\left(\omega\frac{\Delta t}{2}\right)-k^{2}v\,\Delta t\cos\left(\omega\frac{\Delta t}{2}\right)\right)-k_{z}k_{x}\Delta t^{2}C_{0}/\sin\left(\omega\frac{\Delta t}{2}\right)\right\} /4k^{2}k_{z}.\label{eq:C2PSTD}
\end{multline}

Vacuum electromagnetic modes are described by $C_{0}=0,$
\begin{equation}
\sin^{2}\left(\omega\frac{\Delta t}{2}\right)=\left(k\frac{\Delta t}{2}\right)^{2},
\end{equation}
which has as a Courant limit, $\Delta t_{c}=\left(2/\pi\right)\left(\Delta z^{-2}+\Delta x^{-2}\right)^{-\nicefrac{1}{2}}$,
smaller by a factor of $2/\pi$ than the usual FDTD Courant limit.

\begin{figure}
\centering{}\includegraphics[scale=0.95]{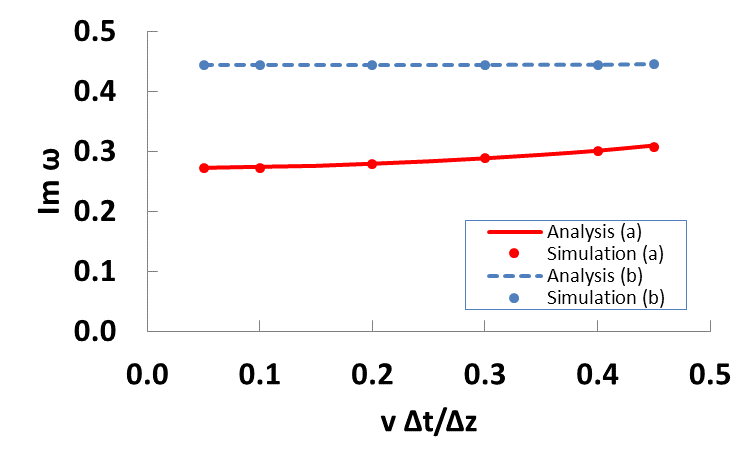}\caption{\label{fig:PSTDlinear-nofiltering}Maximum growth rates for PSTD options
(a) and (b) with linear interpolation and no digital filtering.}
\end{figure}

\begin{figure}
\centering{}\includegraphics[scale=0.95]{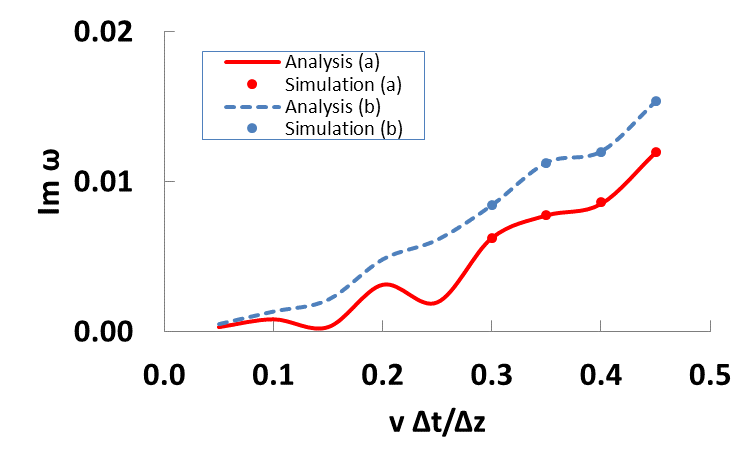}\caption{\label{fig:PSTDcubic-filtering}Maximum growth rates for PSTD options
(a) and (b) with cubic interpolation and digital filtering.}
\end{figure}

PSTD maximum growth rates verses $v\,\Delta t/\Delta z$ are presented
for options (a) and (b) with linear interpolation and no digital filtering
in Fig. \ref{fig:PSTDlinear-nofiltering}. Not surprisingly, these
featureless curves are of the same magnitude as the corresponding
PSATD curves in Fig. \ref{fig:linear-nosmooth} over the same time-step
range. In contrast, maximum growth rates for options (a) and (b) with
cubic interpolation and the digital filtering employed for PSATD,
shown in Fig. \ref{fig:PSTDcubic-filtering}, are an order of magnitude
larger than the corresponding PSATD values at $v\,\Delta t/\Delta z\approx0.4$,
although still very small. This difference results from a narrow region
(about 8 of 4225 k-space modes) at small $k_{x}$ of $m_{z}=0$ non-resonant
instability that does not occur for PSATD. PSTD maximum growth rates
for the same digital filtering and linear interpolation differ only
moderately from the cubic interpolation results.

\section{Simulation results}

Series of two-dimensional simulations of a 100-MeV-class LPA stage
were performed, focusing on plasma wake formation (similar to those
presented in \citep{godfrey2013esirkepov}), using the parameters
given in Table \ref{TableLPA}. With the parameters chosen, dephasing
of the accelerated electron beam and the wake, as well as depletion
of the laser, occur in about 1 mm. However this distance was found
to be too short for any numerical instability to develop with the
pseudo-spectral solvers, and a much longer plasma of 3 cm was used
for the sake of stability analysis. The velocity of the wake in the
plasma corresponds to $\gamma\simeq13.2$, and the simulations were
performed in a boosted frame of $\gamma_{f}=13.$

Reference simulations were run in two dimensions for conditions where
no instability developed, and the final total field energy $W_{f0}$
was recorded as a reference value in each case. Runs then were conducted
for the PSATD and PSTD solvers, using the Esirkepovk current deposition
options (a) and (c) for PSATD, as well as option (a) for PSTD. The
final energy $W_{f}$ was recorded and divided by the reference energy
$W_{f0}$. The ratio $W_{f}/W_{f0}$ is plotted versus time-step in
Fig. \ref{fig:WARP_linear} from simulations using the PSATD solver
with linear current deposition and 4 passes of bilinear smoothing
plus compensation of both current and interpolated fields. (This is
equivalent in the linear regime to the filtering described in previous
Sections.) Following theoretical predictions, option (a) exhibits
no instability for $v\triangle t/\triangle z\lesssim0.3$, $v\triangle t/\triangle z\approx0.5$
and $v\triangle t/\triangle z=1$, and option (c) exhibits an additional
null at $v\triangle t/\triangle z=2$. Fig. \ref{fig:WARP_cubic}
shows results using cubic current deposition, where the PSATD and
PSTD instabilities are contrasted to those of the FDTD Cole-Karkkainnen
(CK) solver with Galerkin or uniform field interpolation. Still in
agreement with theoretical predictions, the PSATD solver is shown
to be stable over a wide range of time-steps for $v\triangle t/\triangle z\lesssim1.2$
with option (a) and even as wide as $v\triangle t/\triangle z\lesssim2.1$
with option (c). The PSTD solver also exhibits good stability but
only on the more restricted $v\triangle t/\triangle z\lesssim0.45$,
owing to its constraining Courant limit.

Note that conducting the time-step sweeps described in this section
would have been prohibitively expensive for the $\gamma=130$ employed
elsewhere in this article. The smaller $\gamma=13$ used here increases
the option (c) growth rates at larger time-steps, as well as the Uniform-CK
growth rates in the vicinity of $v\triangle t/\triangle z=0.5$.

\begin{table}[htd]
\caption{List of parameters for simulations of wake propagation in a LPA stage.}

\begin{centering}
\begin{tabular}{lcc}
\hline 
plasma density on axis  & $n_{e}$  & $10^{19}$~cm$^{-3}$\tabularnewline
plasma longitudinal profile  &  & flat\tabularnewline
plasma length  & $L_{p}$  & $3$ cm\tabularnewline
plasma entrance ramp profile  &  & half sine\tabularnewline
plasma entrance ramp length  &  & $20$ $\mu$m\tabularnewline
\hline 
laser profile  &  & $a_{0}\exp\left(-r^{2}/2\sigma^{2}\right)\sin\left(\pi z/3L\right)$\tabularnewline
normalized vector potential  & $a_{0}$  & $1$\tabularnewline
laser wavelength  & $\lambda$  & $0.8$ $\mu$m\tabularnewline
laser spot size (RMS)  & $\sigma$  & $8.91$ $\mu$m\tabularnewline
laser length (HWHM)  & $L$  & $3.36$ $\mu$m\tabularnewline
normalized laser spot size  & $k_{p}\sigma$  & $5.3$\tabularnewline
normalized laser length  & $k_{p}L$  & $2$\tabularnewline
\hline 
cell size in x  & $\Delta x$  & $\lambda/20$\tabularnewline
cell size in z  & $\Delta z$  & $\lambda/20$\tabularnewline
\# of plasma macro-particles/cell  &  & 4 electrons + 4 protons\tabularnewline
\hline 
\end{tabular}
\par\end{centering}

\label{TableLPA} 
\end{table}

\section{Conclusions}

\begin{figure}
\begin{centering}
\includegraphics[scale=0.95]{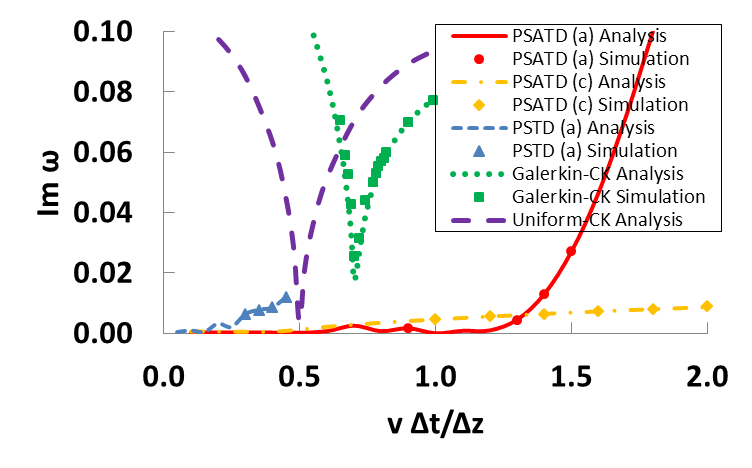}
\par\end{centering}

\caption{\label{fig:Comparison}Maximum growth rates for PSATD (a), PSATD (c),
PSTD (a), Galerkin-CK, and Uniform-CK with cubic interpolation and
digital filtering.}
\end{figure}

The numerical stability properties of multidimensional PIC codes employing
the PSATD electromagnetic field algorithm, combined with either the
Esirkepovk or the conventional current deposition algorithm, have
been derived. Overall, the numerical Cherenkov instability growth
rates for the various versions of the PSATD algorithm are comparable
with those of FDTD algorithms. However, when cubic interpolation and
short wavelength digital filtering also are employed, at least two
versions of the PSATD algorithm exhibits excellent stability over
a wide range of time-steps. For comparison purposes, stability properties
of the more commonly used PSTD electromagnetic field algorithm also
were determined. Fig. \ref{fig:Comparison} compares growth rates
for the most stable versions of these two algorithms (options (a)
and (c) for PSATD and option (a) for PSTD, as defined in Sec. 4 and
again in Table (\ref{Tableoption}).) with the growth rates of the
Galerkin and Uniform versions of the Cole-Karkkainnen \citep{ColeIEEE1997,ColeIEEE2002,KarkICAP06}
FDTD algorithms (coupled with the Esirkepov algorithm), studied in
\citep{godfrey2013esirkepov}. The PSATD options (a) and (c) exhibit
clearly superior stability behavior, with (a) modestly better at smaller
time-steps and (c) substantially better at larger time-steps. Although
the PSTD algorithm also exhibits small growth rates, its range of
time-steps is limited by its relatively small Courant condition. The
two FDTD algorithms have small instability growth rates only over
narrow time-step bands. These findings are corroborated by WARP simulation
results in Fig. \ref{fig:WARP_cubic}.

\section{Acknowledgment}

We thank David Grote for support with the code WARP. This work was
supported in part by the Director, Office of Science, Office of High
Energy Physics, U.S. Dept. of Energy under Contract No. DE-AC02-05CH11231
and the US-DOE SciDAC ComPASS collaboration, and used resources of
the National Energy Research Scientific Computing Center.

\section*{References}

\bibliographystyle{elsarticle-num}


\clearpage{}

\clearpage{}

\clearpage{}

\clearpage{}

\clearpage{}

\clearpage{}

\clearpage{}

\clearpage{}

\clearpage{}

\clearpage{}
\clearpage{}

\clearpage{}

\clearpage{}

\begin{center}
\clearpage{}
\begin{figure}
\centering{}\includegraphics[scale=0.92]{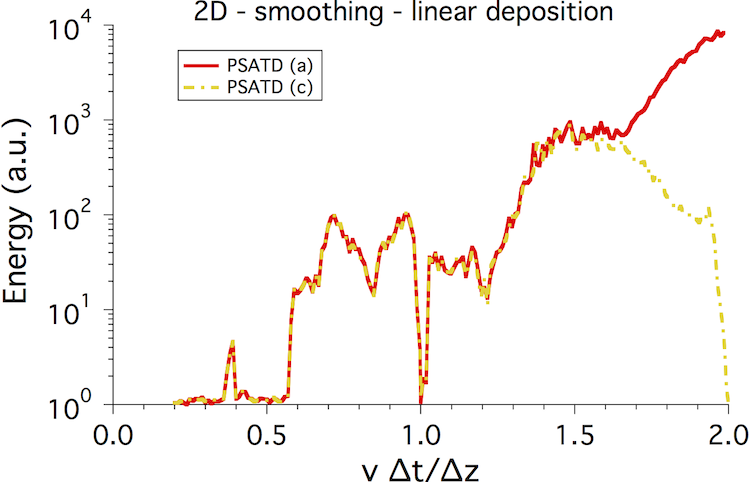}\caption{\label{fig:WARP_linear}Field energy relative to stable reference
level vs $v\Delta t/\Delta z$ from two-dimensional WARP LPA simulations
at $\gamma$ = 13, using the PSATD solver with Esirkepovk current
deposition options (a) and (c), four passes of bilinear plus one compensation
step filtering on both current and gathered fields, and linear interpolation.}
\end{figure}

\par\end{center}

\begin{center}
\clearpage{}
\begin{figure}
\centering{}\includegraphics[scale=0.92]{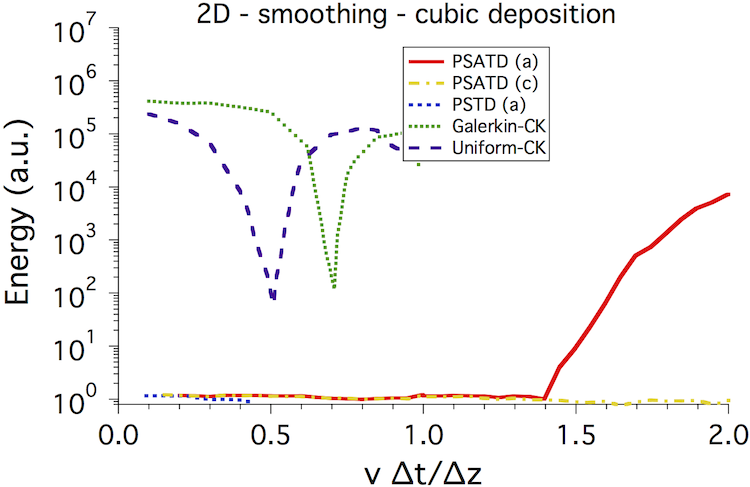}\caption{\label{fig:WARP_cubic}Field energy relative to stable reference level
vs $v\Delta t/\Delta z$ from two-dimensional WARP LPA simulations
at $\gamma$ = 13, using the PSATD or PSTD solvers with Esirkepovk
current deposition options (a) and (c), four passes of bilinear plus
one compensation step filtering on both current and gathered fields,
and cubic interpolation. Results are contrasted to simulations using
the CK solver with Galerking or Uniform field gather, same filtering
and cubic interpolation.}
\end{figure}

\par\end{center}

\clearpage{}
\clearpage{}This document was prepared as an account of work sponsored
in part by the United States Government. While this document is believed
to contain correct information, neither the United States Government
nor any agency thereof, nor The Regents of the University of California,
nor any of their employees, nor the authors makes any warranty, express
or implied, or assumes any legal responsibility for the accuracy,
completeness, or usefulness of any information, apparatus, product,
or process disclosed, or represents that its use would not infringe
privately owned rights. Reference herein to any specific commercial
product, process, or service by its trade name, trademark, manufacturer,
or otherwise, does not necessarily constitute or imply its endorsement,
recommendation, or favoring by the United States Government or any
agency thereof, or The Regents of the University of California. The
views and opinions of authors expressed herein do not necessarily
state or reflect those of the United States Government or any agency
thereof or The Regents of the University of California. 
\end{document}